\documentclass{ar-1col-S2O}
\usepackage[numbers]{natbib}
\usepackage{amsmath}
\usepackage{url}
\usepackage{hyperref}
\usepackage{gensymb}
\usepackage{placeins}   
\usepackage{dblfloatfix} 
\usepackage[utf8]{inputenc}
\jname{Xxxx. Xxx. Xxx. Xxx.}
\jvol{AA}
\jyear{YYYY}
\doi{10.1146/((please add article doi))}

\begin{document}

\markboth{Author et al.}{Accelerator-Based Neutrino Beams}

\title{Accelerator-Based Neutrino Beams}

\author{Laura Fields,$^1$ and Sudeshna Ganguly$^2$
\affil{$^1$Department of Physics and Astronomy, University of Notre Dame, Notre Dame, USA, 46556; email: lfields2@nd.edu}
\affil{$^2$Target Systems Department, Fermi National Accelerator Laboratory, Batavia, USA, 60510}
}

\begin{abstract}
Over the past six decades, accelerator-based neutrino beams have revolutionized particle physics.  Neutrinos created with accelerators have been used to discover the muon neutrino and tau neutrinos was discovered and to confirm the existence of neutrino oscillations. More recently, long-baseline experiments have offered the first experimental hint of CP violation in the neutrino sector. Building and operating such beams is an enormous technical challenge, yet they remain our most versatile tool for studying neutrinos. With new experiments such as DUNE and Hyper-Kamiokande, and ideas such as neutrino factories, the next generation of beams will address open questions about neutrino mass ordering, CP violation, and possible physics beyond the standard model.
\end{abstract}

\maketitle

\tableofcontents


\FloatBarrier
\section{INTRODUCTION}

The universe has provided scientists with copious natural sources of neutrinos, including the Sun, cosmic rays, supernovae, and the decay of element isotopes.
However, there is only one source that gives scientists a modicum of control over the number, flavor, and energy spectrum of neutrinos produced: particle accelerators.  But even here, it would be an exaggeration to say that neutrinos can be controlled. Their unique properties, such as their neutral charge and tiny interaction probability, make creating usable sources of neutrinos an immense challenge.  Modern neutrino beams are engineering marvels that must withstand some of the most radioactive environments on the planet.  

In spite of these challenges, accelerator-based neutrino beams have provided many scientific discoveries, from the first observations of muon neutrinos at Brookhaven National Laboratory (BNL) in 1962~\cite{Danby:1962nd} to the recent evidence for Charge-Parity (CP) violation at the T2K experiment (J-PARC Neutrino Facility) in Japan~\cite{T2K2020}.  And they are poised to continue to deliver science for decades to come.  In this article, we review the current status of accelerator-based neutrino beams, focusing primarily on conventional ``horn-focused" neutrino beams. In particular, we leave the subjects of neutrino beams from Spallation neutron and collider sources, rich topics in their own right, to future review articles.  The remainder of this section briefly reviews the physics goals of the modern accelerator-based neutrino program.  In Section ~\ref{sec:components}, we discuss the key components of conventional neutrino beams.  In Section~\ref{sec:beamsim} we outline the challenges associated with the precise simulations that are necessary for scientific discovery.  Finally, we review past and currently running beams in Section~\ref{sec:operating}, as well as prospects for the future, including some less-conventional but exciting possibilities. 
%

\subsection{Physics Goals}
Although neutrinos were initially discovered using a nuclear reactor source at the Savannah River Nuclear Plant~\cite{Cowan1956}, accelerator-based sources quickly became critical tools for measuring the properties of neutrinos.  The driving physics goal of the recent accelerator-based program has been the measurement of neutrino oscillations.  First hinted at by measurements at Ray Davis' experiment~\cite{Davis1968} at the Homestake Laboratory in the 1960's and confirmed by several experiments including SNO~\cite{SNO} and Super-Kamiokande~\cite{SK}, neutrino oscillations are the phenomenon whereby neutrinos produced as one flavor (electron, muon, or tau) oscillate to a different flavor after traveling through space.  Although neutrinos were originally envisioned to be massless in the standard model of Particle Physics, neutrino oscillations are expected behavior if neutrinos have mass, therefore the observation of neutrino oscillations is typically taken to be synonymous with evidence that neutrinos have mass.  

The modern neutrino standard model is described in detail by the Particle Data Group (PDG)~\cite{PDG}.  To briefly summarize, the three neutrino flavor states ($\nu_e$, $\nu_\mu$, and $\nu_\tau$) are mixtures of three mass states ($\nu_1$, $\nu_2$, and $\nu_3$) with the mixing given by the Pontecorvo-Maki-Nakagawa-Sakata (PMNS) matrix:
\begin{equation*}
   \begin{bmatrix}
    \nu_e \\ \nu_\mu \\ \nu_\tau
   \end{bmatrix}
   =
   \begin{bmatrix}
    U_{11} & U_{12} & U_{13} \\ U_{21} & U_{22} & U_{23} \\ U_{31} & U_{32} & U_{33} 
   \end{bmatrix}
   \begin{bmatrix}
    \nu_1 \\ \nu_2 \\ \nu_3 
   \end{bmatrix},
\end{equation*}
The matrix elements $U_{ij}$ can be expressed in terms of three mixing angles ($\theta_{12}$, $\theta_{13}$ and $\theta_{23}$) and one CP-violating phase ($\delta_{CP}$):
   \begin{equation*}
U = 
\begin{bmatrix} 
c_{12} c_{13} & s_{12} c_{13} & s_{13}e^{-i\delta_{CP}} \\
-s_{12} c_{23} - c_{12} s_{13} s_{23} e^{i\delta_{CP}} & c_{12} c_{23} - s_{12} s_{13} s_{23} e^{i\delta_{CP}} & c_{13} s_{23} \\
s_{12} s_{23} - c_{12} s_{13} c_{23} e^{i\delta_{CP}} & -c_{12} s_{23} - s_{12} s_{13} c_{23}e^{i\delta_{CP}} & c_{13} c_{23}
\end{bmatrix},       
   \end{equation*}
where $c_{ij}$ and $s_{ij}$ denote $\cos \theta_{ij}$ and $\sin \theta_{ij}$, respectively.  In addition to these mixing parameters, neutrino oscillations are additionally dependent on the neutrino mass differences squared, typically specified as $\Delta m^2_{21}$ and $\Delta m^2_{32}$, as well as the ratio of the distance traveled by the neutrino to its energy ($L/E_\nu$).  When the neutrino is propagating through matter, forward neutrino scattering can modify the probability of oscillation in what is known as the Mikheyev-Smirnov-Wolfenstein (MSW) or matter effect.  

Neutrino oscillations observed in different sources are sensitive to different mixing parameters. Neutrino fluxes from the Sun and from nuclear reactors undergo oscillations driven by $\theta_{12}$ and $\Delta m^2_{21}$, known as solar parameters, whereas atmospheric and accelerator-based long-baseline oscillation experiments have characteristic $L/E_\nu$ values governed primarily by the atmospheric parameters, $\theta_{23}$ and $\Delta m^2_{23}$. Both types of experiment are somewhat sensitive to the subdominant parameter $\theta_{13}$.  

In the coming years, key goals of accelerator-based experiments will be the measurement of the CP-violating phase $\delta_{CP}$ and the determination of the ordering of the neutrino mass states, the remaining component of which is the measurement of the sign of $\Delta m^2_{32}$.

In addition to these measurements of parameters within the modern standard model, accelerator-based neutrino experiments also search for physics beyond the standard model in a variety of ways.  Several neutrino experiments have made measurements that, taken at face value, would be consistent with additional neutrinos beyond the three described above.  Combining neutrino measurements with collider measurements that have precisely confirmed that there are three weakly interacting neutrinos, these additional neutrinos would participate in neutrino oscillations but would not otherwise interact with standard model particles. These hypothesized neutrinos are known as sterile neutrinos.  Several dedicated experiments have been constructed to search for sterile neutrinos, and searches can also be conducted in long-baseline experiments designed for other purposes.  Accelerator-based experiments also have sensitivity to other potential beyond-the-standard-model signatures such as non-standard neutrino interactions~\cite{Wolfenstein1978, Valle1987, Farzan2018} and non-unitarity of the neutrino mixing matrix~\cite{Antusch2006, FernandezMartinez2016, Blennow2017}.  There is also the possibility that accelerator-based beams designed to produce neutrinos are also sources of new particles. 
In parallel, these experiments have provided important measurements of standard model parameters, including electroweak couplings and neutrino–nucleon cross sections, and continue to refine constraints on fundamental quantities such as the weak mixing angle. Beyond oscillation studies, a rich experimental program is underway to map out neutrino interactions with nuclei across a broad range of energies, an effort that is essential both for reducing systematic uncertainties in oscillation measurements and for advancing our understanding of nuclear structure.

\section{COMPONENTS AND TECHNICAL CONSIDERATIONS}

In this section, we review the basic components of a conventional neutrino beam. 
Throughout this section, the Neutrinos at the Main Injector (NuMI) beamline and the Long-Baseline Neutrino Facility (LBNF) are used as representative examples, but the design principles and beamline components discussed here are generally applicable to modern accelerator-based neutrino facilities worldwide.

\begin{figure*} 
\includegraphics[width=\linewidth]{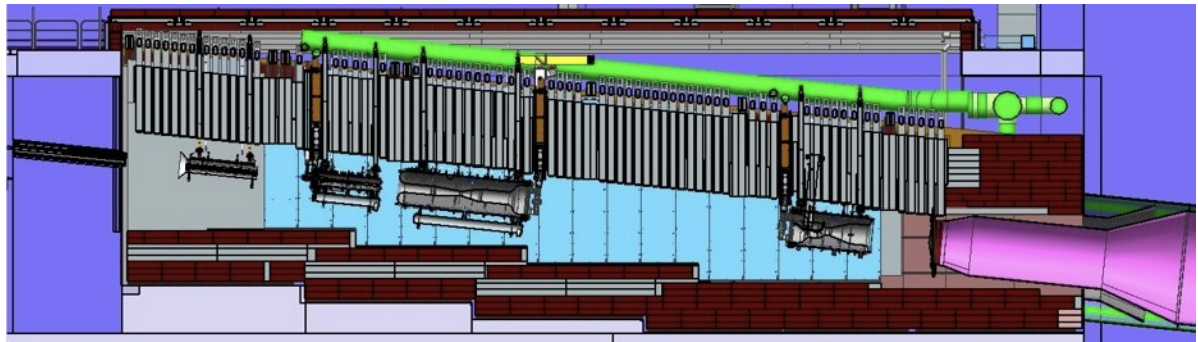}
\caption{\label{fig:schem}
Schematic of the upstream portion of the LBNF neutrino beamline showing the major components. From left to right (in the direction of the beam): the horn-protection baffle, three magnetic focusing horns, and the entrance to the decay pipe. The bulk steel shielding of the target chase is shown in gray and brown, while the water-cooling panels are depicted in light blue. The core elements involved in focusing and transporting the secondary particles toward the decay pipe are shown in this detailed layout. Image courtesy of Fermilab.}
\end{figure*}

As an example, Figure \ref{fig:schem} shows the key components of the LBNF neutrino beamline~\cite{lbnf-tdr}. From left to right (in the beam direction), the schematic shows the horn-protection baffle, three magnetic focusing horns, and the entrance to the decay pipe, all enclosed within heavy steel shielding. 

\label{sec:components}
\subsection{Accelerator Inputs}

The first step in producing a conventional neutrino beam is the delivery of an intense, high-energy proton beam to the target. The target for the primary proton beam is hidden inside the first horn. The characteristics of this proton beam--its energy, intensity, and time structure--directly determine the flux and spectrum of neutrinos that can ultimately be produced. Achieving the required intensities typically involves a sequence of accelerators, beginning with a linear accelerator and followed by one or more circular accelerators that bring the protons to GeV--TeV energies. Since acceleration is usually provided by radiofrequency (RF) cavities, the extracted proton beam carries a time structure set by the RF systems and the layout of the acceleration complex. For example, the NuMI beam at Fermilab is delivered from the Main Injector~\cite{Adamson2016} in $\sim$10~$\mu$s spills, arriving approximately once per second. Each spill contains up to six 1.6~$\mu$s batches; in this context a batch refers to the 84 RF buckets (bunches of protons) that are extracted from the Fermilab Booster in a single cycle.
Within each batch, the protons are grouped into RF bunches at the 53 MHz frequency of the Main Injector, corresponding to a spacing of about 19 ns between bunches.
Once protons are extracted from the parent accelerator, they are directed to the neutrino beamline via a dedicated transfer line. Magnets guide and shape the beam with high precision, making sure that the spatial profile matches the requirements of the target and the focusing system. In circular accelerators, extraction can take place in a single turn (fast extraction) or over extended time intervals (slow extraction). Neutrino facilities generally favor fast extraction to deliver the beam in a short and intense burst, maximizing secondary particle production and capture. The kicker magnets in NuMI provide a rapid horizontal deflection, followed by Lambertson magnets that bend the beam vertically away from the Main Injector. The extracted beam is then steered and focused by dipole and quadrupole magnets---collectively referred to as lattice optics---before striking the target with the required spot size and angle.

\FloatBarrier
\subsection{Targets}
The target is located at the heart of a neutrino beamline. It is where the intense primary proton beam is converted into the mesons that ultimately decay to neutrinos. Targets come in all shapes and sizes but are typically a few nuclear interaction lengths long (e.g., $\sim$1 m for graphite), allowing a large fraction of the primary beam to interact, and narrow ($\mathcal{O}(1~\mathrm{cm})$), minimizing re-interactions of the produced particles. The targets can be cylindrical (for example, T2K) or rectangular (for example, NuMI), and can be continuous or segmented. 


Targets for neutrino beams have been made from a variety of materials. In modern designs, graphite is the most commonly used material because it combines a low thermal expansion coefficient with reasonably good thermal conductivity. The design of targets must take into account a number of competing considerations.  

{\bf Meson production.} Physics goals define the optimal neutrino flavor and energy range, which in turn set requirements on the target geometry and material. These physics needs must be balanced against the engineering constraints described below. 

{\bf Thermal considerations.} Much of the primary beam energy is deposited as heat, and targets routinely operate at hundreds of degrees Celsius. 
Due to the pulsed structure of neutrino beams, the target is subjected to repeated thermal cycles. Active cooling is necessary to prevent overheating and cracking of the target.

{\bf Radiation damage.} 
Neutrino beam targets suffer radiation conditions that are orders of magnitude more powerful than those present in nuclear reactors, because multi-megawatt proton beams deposit their energy in a very small volume over microsecond timescales, creating extreme energy densities and material damage.
The resulting damage changes the bulk properties of the target material, including its hardness, shape, and thermal properties~\cite{Mokhov2005,Davenne2012}.  
In accidents, targets have exhibited localized twisting and swelling, whereas under sustained irradiation, the dominant effects are microstructural changes in the crystalline lattice that progressively alter thermomechanical response. 

{\bf Integration with the focusing system.} To capture as many low-energy pions as possible, part of the target is often placed inside the first focusing horn. Although this boosts the useful flux, it leaves less room for mechanical supports and makes target replacement more difficult.  

{\bf Remote handling.} The targets must be changed regularly due to radiation damage, manufacturing faults, or accidents during operation. For this reason, they are designed to be removed and replaced remotely. Once removed, the targets are far too radioactive to handle directly and must be stored in shielded facilities until their activity drops to safe levels for disposal.

{\bf Baffles} Targets are often preceded by robust graphite or carbon–carbon baffles that intercept misdirected protons before they strike downstream components. The simple geometry of the primary beam tube and baffle aperture defines the maximum excursion of any mis-steered beam: once the beam moves beyond this envelope, it is absorbed. In practice, the baffle (and, where present, a small upstream ``bafflette'' integrated with the target) protects not only the horns but essentially all downstream components, a fact explicitly considered in accident analyses of the absorber and chase.

\FloatBarrier
\subsection{Focusing Systems}
Magnetic focusing horns were first proposed by Simon van der Meer in 1961~\cite{vanderMeer:1961}. 
Van der Meer compared his invention to a conical reflecting surface that focuses light emitted from within the cone. 
Each time the light reflects from the inner surface, it becomes more parallel until it is straight enough to exit the cone. 
In a magnetic horn, two nested conductors carry a pulsed current: it flows along the inner conductor in one direction and returns along the outer conductor in the opposite direction. 
By Ampère's law, this current produces a magnetic field only in the region between the conductors, with field lines wrapping in circles around the central axis of the horn, that axis being parallel to the incoming proton beam. 
This field bends the trajectories of charged pions and kaons, focusing those that travel in the forward direction into a parallel beam suitable for decay into neutrinos. 
Horns act only on charged \emph{secondaries} (pions/kaons); the primary proton beam is not focused by the horns.

An example is NuMI Horn 1, shown in Fig.~\ref{fig:horncurrent}.
The horn is made of coaxial aluminum conductors pulsed at about 200~kA (studies occasionally to $\sim$210–220~kA); by contrast, LBNF horns are designed for $\sim$300~kA operation, producing toroidal fields of up to 3~T. 

This field bends the trajectories of (primarily) charged pions and kaons from the target, focusing one charge-sign into the decay pipe and bending the other charge-sign away.  Positively charged pions decay to a positive muon and a neutrino, while negatively charged pions decay to a negative muon and an antineutrino.  Therefore, the direction of the current in the horn can be used to produce a concentration of neutrinos or antineutrinos.

When the current flow is in the direction labeled forward horn current (FHC) mode, positively charged pions are focused, while negatively charged pions are defocused; reversing the polarity of the power supply produces an antineutrino beam in the Reverse Horn Current (RHC) mode. 
Both NuMI horns have parabolic profiles that act as chromatic lenses, focusing particles in a specific momentum range. 
A curved conductor closes the circuit at the upstream end, while a stripline supplies current at the downstream end, with both horns connected electrically in series. 
Because meson decays largely produce muon neutrinos, the resulting beam is primarily muon-neutrino in composition.
Figure~\ref{fig:horncurrent} shows the current flow and pion trajectories in neutrino mode. 
High-energy pions passing near the horn neck, where the field is strongest, are not fully deflected because of their large momentum. Some of these poorly focused pions have the wrong charge and contribute to wrong-sign contamination, while additional electron-neutrino backgrounds come from kaon and muon decays.
For NuMI and LBNF horns, the outer conductor is anodized to provide an insulating barrier to prevent drift currents from distorting the torroidal magnetic field.

\begin{figure*}
\includegraphics[width=\linewidth]{Figures/horn1currentflow.png}
\caption{\label{fig:horncurrent}
NuMI Horn 1, with a red arrow indicating the current flow direction in FHC mode.
Trajectories of pions with different initial momenta. Red: 5\,GeV/$c$, green: 10\,GeV/$c$, blue: 20\,GeV/$c$, black: A 10\,GeV/$c$ pion with opposite charge (``wrong-sign'')is defocused by the horn running in  the FHC mode. The beam travels from left (upstream) to right (downstream) in this figure. Figure adapted from Ref.~\cite{q6l6-wywy}.
}
\end{figure*}

During beam operation, horns experience heat transfers due to both energy deposition of beam particles and Joule heating from current circulating through the horn.  Because of its proximity to the beam and greater current densities, the inner conductor is subject to greater heat stresses than the outer conductor. Modern horns are cooled by high-pressure water sprays that form a thin film on the conductor surfaces for rapid heat removal; in some designs, the conductors are nickel-coated to reduce corrosion.

\FloatBarrier
\subsection{Decay Volumes, Beam Dumps, and Shielding}
Downstream of the target and the focusing horns, the next major component of a neutrino beamline is the decay volume, where secondary mesons decay into neutrinos.
Earlier facilities used shorter decay pipes, but today these decay regions tend to be several meters across and extend hundreds of meters. 
The choice of length represents a balance: a longer decay volume increases the probability that pions and kaons will decay in flight, enhancing the neutrino flux, but also requires more extensive civil construction and shielding. 
Examples include the 675~m-long, helium-filled decay pipe in NuMI and the 96~m-long, helium-filled decay pipe used in T2K.  In NuMI, the higher-momentum pions ($\sim$10~GeV/$c$) require a longer decay volume, whereas the shorter decay pipe used in T2K (optimized for lower-momentum pions around $\sim$2~GeV/$c$) reflects this design trade-off.
 Future facilities such as LBNF employ lengths of order 200~m with optimized geometry and shielding to maximize flux while managing cost and radiological concerns.  

In the decay region, only a few percent of the initial proton beam energy is ultimately carried away by neutrinos; the remainder must be safely absorbed in shielding or beam dumps. 
To protect the surrounding environment, including personnel, groundwater, and surrounding infrastructure, thick layers of concrete, steel, and, depending on elevation or depth, soil or rock are placed around the target hall and the decay pipe. 
Although this shielding has little effect on the on-axis neutrino flux reaching detectors, it does prevent hadrons from escaping into the environment and enhances the low-energy decay-at-rest component of the beam. 
Shielding design also strongly influences off-axis fluxes, which are exploited in some experiments.  

At the downstream end of the decay volume sits a beam dump, or hadron absorber, designed to stop nearly all residual particles except neutrinos and high-energy muons. 
The absorber typically consists of a core of water-cooled metals (combinations of aluminum, graphite, and steel have been used) surrounded by additional shielding. 
Because muons are the only charged particles that can penetrate the absorber, muon monitors are usually placed downstream to provide a valuable diagnostic of beam alignment and intensity.  
Further muon shielding, often consisting of rock or earth berms, ensures that only neutrinos emerge from the facility.  

\FloatBarrier
\subsection{Beam Instrumentation}
Monitoring the proton beam and the secondary particles at the target is essential for both protecting the beamline and predicting neutrino fluxes. It provides operators with immediate feedback on beam quality and alignment, while providing physicists with information about systematic uncertainties in oscillation measurements.

Several devices are used to track and control the primary proton beam. Beam position monitors, beam profile monitors, and beam current transformers measure beam properties and feed the Machine Protection System (MPS), which can abort to upstream dumps before extraction or inhibit beam within one or two pulses after extraction when interlocks trip. Beam loss monitors (BLMs) principally watch for anomalous proton losses in the beam pipe; special BLMs are temporarily installed for low-intensity scan campaigns and removed afterward. Muon monitors downstream of the absorber provide an indirect measure of neutrino flux and beam–target alignment. Near the target, the thermocouples track the heating from the beam, while alignment systems verify that the protons strike the target within an acceptable range of the intended position. Together, these diagnostics add protection and ensure that the beam is delivered to the target under well-controlled conditions, reducing uncertainties associated with the flux.


Downstream detectors complement these measurements within a narrow tolerance range of the intended position and angle. 
Hadron monitors at the end of the decay pipe sample the charged particles which survive without decaying, and additional detectors behind the beam absorber sample the exiting muons, which act as a proxy for the neutrino flux.
The muon signal changes predictably with horn current or alignment, allowing these detectors to track long-term beam stability~\cite{q6l6-wywy}. At NuMI, three muon monitors located at different shielding depths provide sensitivity to different momentum ranges of the muon and therefore the neutrino beam.

Beam-based alignment adds another layer of precision. After surveying the positions of the target and horns, controlled scans using a low intensity proton beam are performed and correlated with downstream detector signals. This technique refines the absolute positions of the beamline elements relative to the beam instrumentation and ensures that the neutrino beam is properly oriented and characterized~\cite{Zwaska_2006}.

\subsection{Radiation Considerations}

High-power neutrino beams present significant radiation challenges, from secondary showers to long-term activation of beamline components. Three issues dominate: prompt radiation produced during beam operation, the accumulation of residual activation in beamline components, and the risk of activating the surrounding soil, water, and air.  

Neutrons are copiously produced throughout the target station, decay region, and absorber and dominate many shielding and activation considerations. 
Prompt radiation comes from showers of secondary particles created in the target, decay pipe, and absorber. To protect people and nearby equipment, the facilities surrounding these areas are equipped with thick layers of steel, concrete, and earth. The exact design is guided by detailed simulations to keep radiation below regulatory limits, even at multi-megawatt beam powers~\cite{PIPII-SAD, LBNE-design}. Important isotopes for radiation protection include $^{41}$Ar produced in target chase air, $^{22}$Na generated in metallic components, and $^{3}$H (tritium) formed in concrete and other shielding materials. In particular, tritium and sodium-22 are of concern for groundwater and cooling systems and are carefully monitored and contained.

Residual activation is unavoidable whenever the proton beam or secondary showers interact with the hardware. Targets, horns, decay pipes, and absorbers can all become highly radioactive. This makes the use of remote handling systems essential, with strict access controls and continuous dose monitoring during beam-off maintenance operations. 
Protecting the environment is equally important. Cooling water and groundwater must be contained to avoid the spread of activated isotopes. The air in and around target stations can pick up short-lived radionuclides such as $^{11}$C, $^{13}$N, and $^{41}$Ar, which require careful ventilation and monitoring to prevent release to the environment~\cite{NuMI-tritium, Hylen-target-stations}.  


Finally, real-time radiation protection monitors provide the last line of defense for life safety: if levels exceed thresholds, access controls interlock, and the beam is shut off, typically through devices such as Fermilab's ``chipmunk'' ionization chambers. These channels feed the radiation protection system (RPS) first and foremost (life safety) and may also provide secondary input to the machine protection system (MPS). With this layered approach—shielding, environmental controls, RPS interlocks, and MPS trips—modern neutrino facilities achieve high intensity while protecting workers and the environment~\cite{DOE-EA-1267}.


\FloatBarrier
\section{BEAM SIMULATIONS}
\label{sec:beamsim}
\begin{figure*} 
\includegraphics[width=\linewidth]{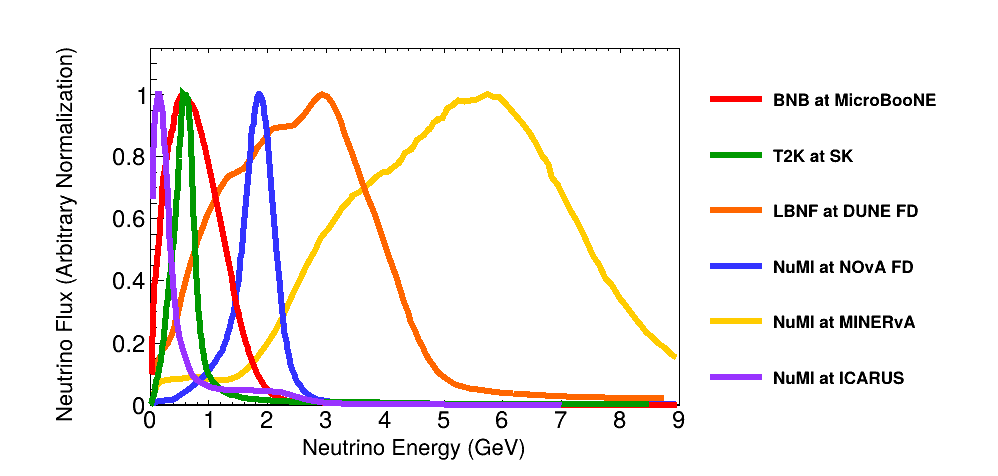}
\caption{\label{fig:flux}
Neutrino flux predictions at a selection of current and future neutrino detectors.}  
\end{figure*}

Figure~\ref{fig:flux} shows representative neutrino flux predictions at a selection of current and future neutrino detectors.

Accelerator-based neutrino experiments measure the energy spectra and other details of neutrino interactions induced by their neutrino beam.  
In order to transform these spectra into meaningful measurements, detailed simulations of the entire experiment, from the neutrino beam to the neutrino detector, are required.  Here we focus on the first element of that -- the neutrino flux simulation.  These are produced using a software toolkit that simulates the interactions of particles with matter, beginning with the primary proton beam interaction with the target and propagating all secondary particles through a detailed material simulation of beamline elements until a hadron decays to a neutrino.  Detailed information about the neutrino's kinematics and its particle ancestry is written out and subsequently fed as input to a neutrino interaction generator.  

Because the hadronic interactions in these simulations involve the non-perturbative regime of QCD, precise predictions of interaction cross sections and final state kinematics are not available.  An extensive effort to measure relevant interactions is currently underway by the EMPHATIC~\cite{emphatic-proposal, emphatic-measurement} and NA61/SHINE experiments~\cite{na61-2010, na61-2018, na61-2025}.  These data will provide more precise flux predictions than are possible with out-of-the-box toolkit predictions and also provide a means of estimating uncertainties on neutrino fluxes, which are also required input to measurements with neutrino beams.  These flux uncertainties currently run between 4 and 50\% depending on the neutrino energy and flavor.  Further constraints on neutrino flux uncertainties using data from neutrino detectors is possible but challenging.  For the remainder of this section, we discuss the elements of precise neutrino flux prediction in more detail.

\subsection{Software Toolkits}
The most commonly used software toolkit for neutrino beam simulations is Geant4~\cite{geant4a,geant4b,geant4c}, a toolkit for simulating the passage of particles through matter that is used for a wide array of applications, including particle, nuclear, medical, and space science.  Geant4 is in active development by a collaboration of scientists, with hadronic physics models and toolkit functionality improvements provided on a regular basis through new code releases.  Users may select from a variety of ``physics lists" (or combinations of hadronic models) based on applications.  It provides a flexible geometry interface in which users can construct material and magnetic models natively in a C++ application or import via, e.g., gdml files~\cite{gdml}.  

Although these features have made Geant4 the tool of choice for modern neutrino beam simulations, two additional software packages have also been used for neutrino flux predictions.  Fluka~\cite{Fluka1,Fluka2} is another general purpose particle transport framework.  It is known for accurate modeling of hadron production, but it has less flexibility in geometric modeling. For this reason, it has been used by several experiments for modeling interactions in neutrino beam targets, the products of which are fed into Geant for subsequent propagation through the beamline.  


MARS~\cite{mars15} is a particle–matter transport and radiation analysis code developed for a long time at Fermilab (originally introduced there by N.~Mokhov). Although used primarily for radiological studies at Fermilab, it can also produce neutrino flux predictions and has been applied to LBNF.

Regardless of the software toolkit, a neutrino beam simulation will typically be configured so that the user can specify beam parameters such as primary beam momentum, spot size, and position on target.  It will also contain a detailed description of any of the geometry of the beamline relevant to neutrino production, including the target, horns, decay volume, and hadron absorber, as well as peripheral details such as support and shielding materials.  
The most common output of a beam simulation is a list of neutrinos with kinematic information about both the neutrino, its immediate parent, and sometimes full ancestry history.  Ancestry history can be useful when correcting or assessing systematic uncertainties in hadron production models.  Ancestry kinematics can also be used to ``re-decay" the hadron, forcing it to point in the direction of a detector of interest, which provides a substantial reduction in the computing resources necessary to simulate a neutrino flux at a specific detector.  For Fermilab neutrino beams, the standard output format is defined by the ``DK2NU" package~\cite{Dk2nu}, which provides a structure for storing kinematic and ancestor information about decay.  Although neutrino kinematics is the primary output of a neutrino beam simulation, a variety of other information can be extracted from these simulations for various purposes, such as radiological and beam monitoring studies.  

\subsection{Hadron Production Measurements}

In order to propagate particles through the beamline, Geant4, FLUKA, and MARS all implement models of elastic and inelastic scattering of particles including protons, pions, and kaons.  Although these models have all been tuned to relevant hadron production datasets, comparisons of predictions with recent hadron production datasets indicate residual discrepancies between data and models on the order of tens of percent.  To improve the accuracy of neutrino flux predictions, experiments often re-tune simulations using hadron production data.  

In modern hadron production experiments, there are two broad categories of measurements.  The first category is thin-target measurements, wherein a beam of known momentum, charge, and particle species is collided with a fixed target that comprises only a small fraction of a nuclear interaction length. Differential cross sections describing the probability that the incident particle interacts with a target nucleus are extracted as a function of final-state particle kinematics. These measurements can then be used to tune hadron production models in toolkits such as Geant4 and/or to correct neutrino flux predictions at analysis time through reweighting.  Datasets from experiments including NA49~\cite{NA49}, HARP~\cite{HARP}, and MIPP~\cite{MIPPTT} are currently in use by neutrino experiments.  The NA61/SHINE experiment at CERN and the EMPHATIC experiment at Fermilab are now operating and aim to fill in areas of unmeasured phase space. 

The second category of target measurements is ``replica target" wherein precise replicas of actual neutrino beam targets are placed in hadron production experiments, impinged with a beam similar to (but much lower intensity) than in the neutrino beam, and multiplicities (i.e., yields as a function of momentum and angle) of hadrons leaving the target are measured.  This technique has been used successfully for the BNB beam~\cite{HARP:2007dqt} and T2K beam~\cite{T2KHP}, where uncertainties were lowered to approximately 5\% in the beam focusing peak using ``replica target'' measurements at NA61/SHINE. ``Replica target" measurements for NuMI, DUNE and further T2K measurements are currently underway at NA61/SHINE.  The EMPHATIC experiment at Fermilab aims to take the ``replica target''
 concept a step further, making measurements with not only a ``replica target''
 but also a neutrino focusing horn. 
The first measurements with an unpowered NuMI Horn 1 are planned for 2026.  

Once hadron production data are available, incorporation of these data into flux predictions remains a formidable challenge.  If``replica target'' data are available, this is usually the preferred option.  But even in cases where ``replica target'' data are available, many beamline interactions will not be covered by those data and must use thin-target data if available.  Data are collected at one or a few incident energies and extrapolated to other incident energies using the Feynman scaling~\cite{Feynman}.  Extrapolation from data on one incident material to other materials has also been implemented.  Significant efforts in this area have been made by the T2K collaboration~\cite{T2KFlux} and by the PPFX collaboration~\cite{PPFX} for the NUMI beam.  

\subsection{Neutrino Flux Uncertainties}

\begin{figure*} 
\includegraphics[width=\linewidth]{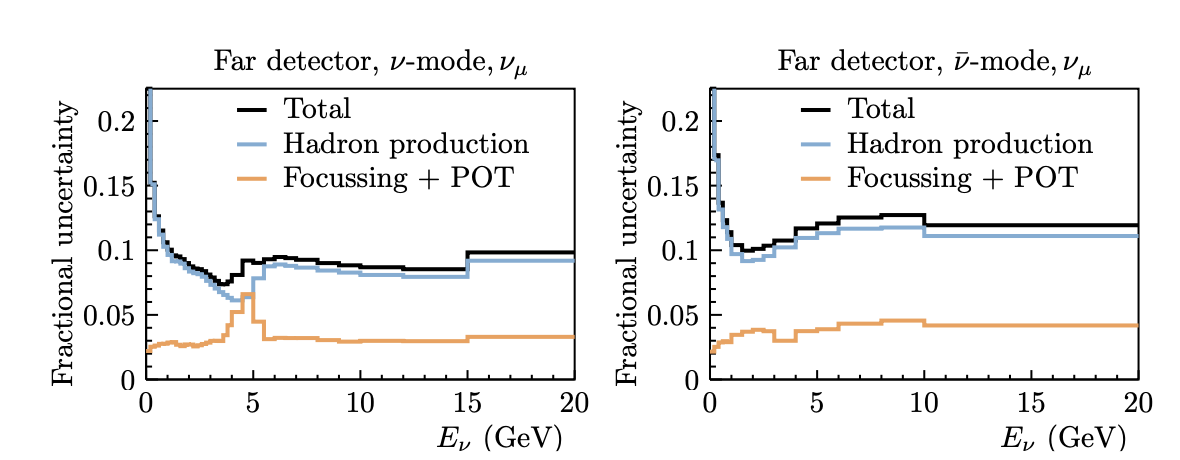}
\includegraphics[width=\linewidth]{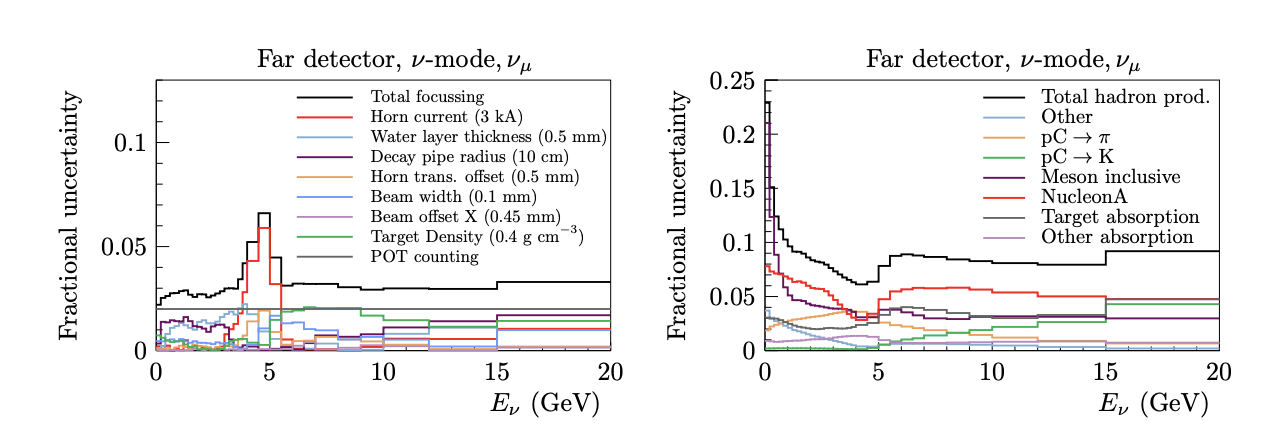}
\caption{\label{fig:flux_uncertainty}
Muon neutrino flux uncertainties at the DUNE far detector in neutrino mode (top left) and antineutrino mode (top right).  Also shown are breakdowns of the hadron production (bottom left) and focusing uncertainties (bottom right) for muon neutrinos in neutrino mode.  Figure reproduced from ~\cite{DUNETDR} and courtesy Luke Pickering.  }
\end{figure*}

Uncertainties in neutrino fluxes arise from underlying uncertainties in the hadron production models used to simulate the flux, as well as uncertainties due to deviations of operating beams from ideal simulated beams (commonly called focusing uncertainties).  As of this writing, hadron production uncertainties dominate the uncertainty budgets of all operating neutrino beams.  As an example, the uncertainties of the estimated flux at the DUNE far detector are shown in Figure~\ref{fig:flux_uncertainty}.  
Here, hadron production uncertainties are dominant everywhere except in the $\sim5$ GeV region, where focusing errors become large.  This is typical for horn-based neutrino beams, where focusing uncertainties tend to increase at the falling edge of the flux-focusing peak. 
In this region, many neutrino parents are hadrons that pass close to the horn neck, where the magnetic field is strongest, making the resulting flux particularly sensitive to small shifts in horn alignment or current.

Looking in detail at the sources of hadron production uncertainty, DUNE's flux uncertainties are dominated by the category labeled in figure~\ref{fig:flux_uncertainty} as ``Nucleon A''
.  This is the uncertainty due to proton and neutron interactions with nuclei, for which there is no directly applicable external data.  This uncertainty is expected to be reduced in the future by new data from NA61/SHINE and EMPHATIC.  Another large uncertainty is the ``meson inclusive''
 category, which accounts for all uncertainties in modeling interactions where the primary particle is a meson.  This will also be reduced by further hadron production data.

\subsection{In Situ Flux Constraints}

Measurement of neutrino fluxes using neutrino detectors is a challenge because neutrino detectors measure the convolution of neutrino flux and neutrino cross sections, and models of neutrino-nucleus interactions carry similarly large uncertainties to the hadronic interactions that occur in neutrino beams.  At present, the most promising processes for measuring neutrino flux are electroweak processes that can be predicted most precisely.  The MINERvA experiment has used neutrino and antineutrino scattering on electrons~\cite{minerva1,minerva2,minerva3} as well as inverse muon decay~\cite{minerva4} as standard candles for flux prediction.  Both processes are relatively rare and provide limited information about the flux shape in relation to neutrino energy, yet MINERvA has reported a total flux normalization uncertainty of less than 4\%, and the DUNE experiment plans to follow suit.  
In the MINOS experiment, muon-monitor signals were used downstream of the absorber to constrain the overall flux normalization to approximately 15\%, demonstrating the usefulness of beam-derived flux probes for oscillation analysis. Moreover, upcoming NOvA studies will incorporate $\nu_e$ appearance samples to improve flux prediction accuracy by using in situ detector measurements.
\FloatBarrier
\section{OPERATING BEAMS}
\label{sec:operating}

\begin{figure*}[!t]
\includegraphics[width=0.9\linewidth]{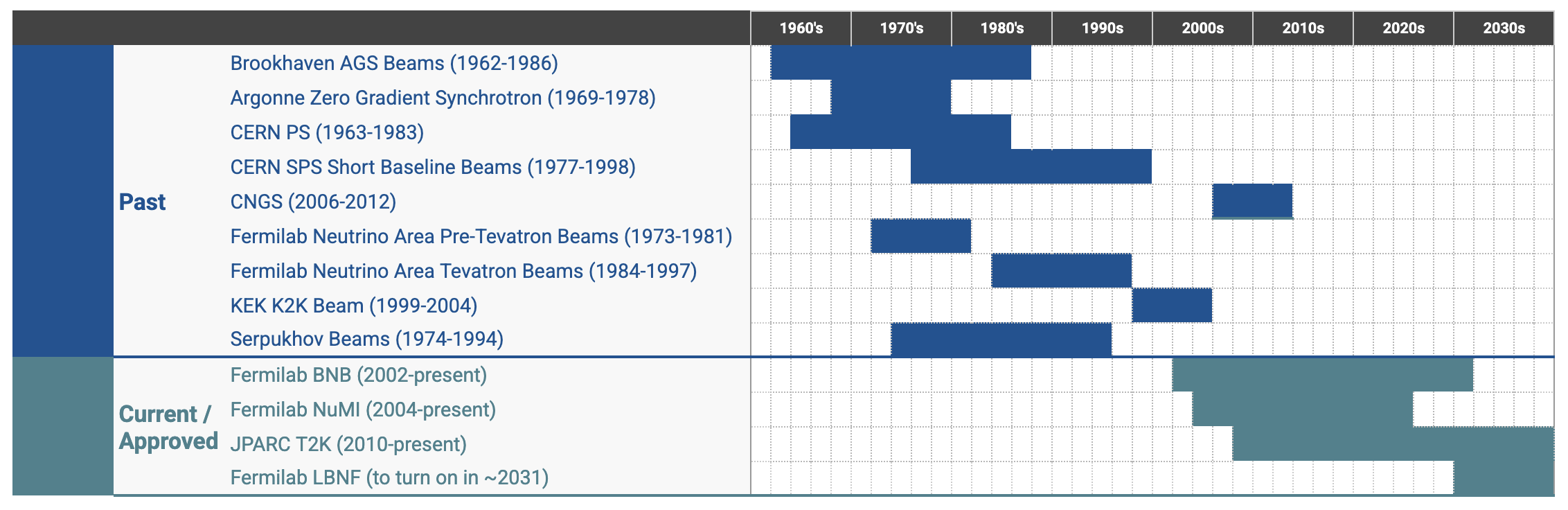}
\caption{\label{fig:timeline}
Time line of past, present and future conventional neutrino sources. }
\end{figure*}
\FloatBarrier
Since the mid-20th century, conventional neutrino beams have been operated in laboratories in the United States, Japan, and Russia.  In this section, we catalog past and present beams and discuss future possibilities.  We note that he development and operation of these facilities have benefited greatly from international coordination and knowledge sharing through the International Workshop on Neutrino Beams and  Instrumentation (NBI) series, which has provided a long-term forum for collaboration among laboratories worldwide pursuing high-power neutrino beam technologies.  

\subsection{Historical Beams}

In 1962, BNL became the site of the first accelerator-based neutrino beam~\cite{Danby:1962}.  Leon Lederman, Melvin Schwartz, and Jack Steinberger used 15 GeV protons from the Alternating Gradient Synchrotron (AGS), a beryllium target, and a 21-m decay volume to produce muon neutrinos with a mean energy of approximately 5 GeV. A downstream detector observed interactions attributable to muon neutrinos, confirming the existence of the muon neutrino and establishing that it is distinct from the electron neutrino.  This pivotal result confirmed that there were multiple flavors of neutrinos.  The experiment's success earned a Nobel Prize in 1988 and was a blueprint for subsequent neutrino beamline design.

BNL’s AGS continued to contribute to neutrino physics for decades.  Focusing horns were added to the original AGS neutrino beam concept in order to provide wide-band and narrow-band beams in the few GeV energy regime for experiments including E734~\cite{E734} and E776~\cite{E776}, which investigated neutrino scattering and searched for oscillation signals.  These beamlines provided key input into the early discussions of neutrino mass and mixing, setting the stage for future long-baseline experiments.

Argonne National Laboratory (ANL) operated neutrino beams using 12.4 GeV/c protons from the Zero Gradient Synchrotron from 1969 to 1978 ~\cite{CourierANL}. These were directed into the ANL 12-foot diameter bubble chamber – the largest bubble chamber in the world at the time.   The Argonne program yielded the first precision measurements of neutrino quasielastic scattering and Delta(1232) resonance production on hydrogen and deuterium.

Soon after the discovery of muon neutrinos, CERN began using Proton Synchrotron (PS) as a neutrino source~\cite{CourierNeutrinos}.  This started in 1963 with the first-ever attempt to use Simon van der Meer's focusing horn concept to focus hadrons and increase neutrino fluxes.   
In this case, protons with momentum of 24.8 GeV were directed onto a 25 cm long copper rod producing pions and kaons that decayed in a 25 m decay volume. A series of bubble chamber experiments, beginning with a 500-L CF$_3$Br chamber, recorded neutrino interactions. The neutrino beam had a mean energy of approximately 1.5 GeV~\cite{CERNNeutrinoExperiment}.
This beamline went through a series of upgrades and was eventually the first home of Gargamelle, a storied 12 m$^3$ Freon bubble chamber that made the first observation of neutral-current neutrino interactions~\cite{Gargamelle}.   In 1977, CERN neutrino operations moved to the new Super Proton Synchrotron (SPS)~\cite{CERN-SPS}, which provided 350 GeV protons and a 290 m decay volume.  Gargamelle was briefly moved to SPS before being decommissioned. The SPS beamline was also home to the Big European Bubble Chamber (BEBC), CERN-Dortmund-Heidelberg-Saclay (CHDS) detector, and CHARM.  The PS was briefly revived as a neutrino source in 1983 so that BEBC, CHDS, and CHARM could search for neutrino oscillations, although none were observed~\cite{CHARM}.

The most recent CERN neutrino endeavor was the CERN Neutrinos to Gran Sasso (CNGS) project~\cite{CNGS}, operational from 2006 to 2012. It was Europe's first long-baseline high-intensity neutrino beamline, designed to measure the oscillation channel $\nu_\mu \rightarrow \nu_\tau$~\cite{opera2015} using the OPERA detector.  The ICARUS detector also operated on the CNGS beamline. The beam was generated using 450 GeV protons, a graphite target, magnetic horns, and a 994~m decay tunnel. The resulting $\nu_\mu$ beam traveled 730 km through the Earth to the Gran Sasso National Laboratory (LNGS) in Italy.  In 2010, OPERA reported the observation of its first candidate event $\nu_\tau$, providing direct evidence of the $\nu_\mu \rightarrow \nu_\tau$ oscillation and supporting the three-flavor oscillation framework.  

Fermilab began its extremely successful neutrino program in the 1970s with a first series of neutrino beams using 300-450 GeV protons~\cite{Kopp}. 
In its initial configuration, the beamline used a bare target with no magnetic focusing; the secondary mesons then entered a cylindrical 350 m long decay pipe ~\cite{HPWF1}.
The beam was later upgraded to include horn focusing.  In this era, Fermilab also created the first ``dichromatic" neutrino beam~\cite{dichromatic} formed using quadrupole magnets downstream of the target to select pions and kaons in a narrow momentum range.  Subsequent decay of the pions and kaons produces a double-peaked (hence the name dichromatic) energy spectrum.     
These various beams were observed by experiments including HPWF~\cite{HPWF1}, CITF~\cite{CITF},  CITFR~\cite{CITFR}, and the 15' Bubble chamber~\cite{BubbleChamberReflections}, resulting in a rich array of physics measurements including confirmation of the existence of neutral-current neutrino interactions first discovered by Gargamelle, many studies of neutrino interactions, and the discovery of neutrino tridents~\cite{Morfin}, a rare weak-interaction process in which a neutrino, scattering off the Coulomb field of a heavy nucleus, produces a charged-lepton pair.     

In the early 1980s, the Fermilab neutrino area was modified to accommodate protons from the Tevatron collider.  The first such beam was the Fermilab Tevatron Quadrupole Triplet beam which collided 800 GeV protons with a BeO target producing hadrons that were selected with quadrupole magnets.  This beam was used by the CCFR, FMMF~\cite{FMMF} collaborations, and the E-632 experiment~\cite{E632PhysRevD}, which used holographic imaging in the 15-ft chamber to measure charm production.  A Tevatron-based beam was later used by the DONUT collaboration in 1997, leading to the discovery of the tau neutrino ~\cite{PhysRevD.78.052002}.  

In Japan, the KEK to Kamioka (K2K) beam operated from 1999 to 2004, sending neutrinos 250 km to the Super-Kamiokande detector. K2K was the first experiment to confirm atmospheric neutrino oscillations using an artificial source, providing critical momentum for future long-range efforts~\cite{k2k2006}.

In Russia, the 70 GeV Serpukhov proton accelerator was used to generate neutrino sources over several decades starting in 1974, when fast-extracted protons were directed onto an one-interaction-length target and focused with a novel focusing system before being directed into a 140 m long decay volume~\cite{Serpukhov}. 
This beam was observed by several detectors, including the ITEP–IHEP spark spectrometer~\cite{ITEP-IHEP-spark}, an upgrade of that detector that added emulsion detectors (IHEP–ITEP–JINR)~\cite{IHEP-ITEP-JINR-emulsion}, and the SKAT bubble chamber~\cite{SKAT-mue-final}, resulting in physics results on neutrino cross sections and tridents.  
In 1987, Serpukhov was upgraded in intensity with the addition of a booster, which in the early 1990s provided beam to a new Tagged Neutrino Facility (TNF)~\cite{TNF-project} that introduced a tagging station between the decay pipe and muon shielding that attempted to measure charged particles from kaon decays.   
In 1993–1994, Serpukhov protons were used in conjunction with an aluminum target and a 12~m decay volume, creating a neutrino beam with a mean energy of 7~GeV to search for sterile neutrinos~\cite{Borisov:1996kv}.

Precision experiments are possible today thanks to these historic neutrino beamlines.
They helped establish the existence of multiple neutrino flavors, neutrino mass, and flavor oscillations. Technological innovations such as magnetic horn focusing, long decay tunnels, and near/far detector configurations became standard features in beamline design.

The design of neutrino beamlines has evolved significantly since the early days of BNL. The basic components—high-energy proton accelerators, meson production targets, magnetic focusing horns, decay tunnels, and shielding—remain largely consistent, but their scale and precision have dramatically improved. Early beamlines had limited focus and control over the resulting neutrino spectrum. 
Multi-horn systems and sophisticated targetry allowed beamlines like NuMI and CNGS to tailor neutrino energy profiles more precisely.
Furthermore, beam monitoring systems have advanced, providing real-time feedback and alignment for long-baseline experiments.

\subsection{Current and Planned Beams}
\label{sec:current}
\subsubsection{Tokai to Kamioka (T2K)}

The Tokai to Kamioka (T2K) experiment in Japan was the first to observe the appearance of electron neutrinos in a primarily muon neutrino beam, providing another direct confirmation of three-flavor oscillations. 
Using the Super-Kamiokande water Cherenkov detector at 295~km as its far detector and the ND280 complex to characterize the unoscillated beam, T2K delivered a sequence of landmark results. 
In 2011 it reported the first hint of $\nu_\mu \!\to\! \nu_e$ appearance, and by 2014 the observation exceeded the $5\sigma$ threshold, firmly establishing the phenomenon and opening the way to precision measurements of leptonic CP violation~\cite{T2K2,T2KObservation2014}.  

The neutrino beam is generated at the Japan Proton Accelerator Research Complex (J-PARC), where 30~GeV protons strike a graphite target to produce pions and kaons. 
These secondary particles are focused by three magnetic horns and decay in a 96~m helium-filled tunnel, yielding a predominantly muon-neutrino beam peaked near 0.6~GeV. 
The beam is directed 2.5$^\circ$ off-axis toward Super-Kamiokande, a geometry that sharpens the spectrum around the first oscillation maximum while suppressing high-energy backgrounds~\cite{T2K}. 
Near detectors at 280~m provide essential flux and cross-section measurements that constrain systematic uncertainties.  

Over time, T2K introduced a series of upgrades to enhance sensitivity to CP violation. 
The horn current was increased from 250~kA to 320~kA to improve focusing~\cite{OyamaJPARCUpgrade}, and the ND280 complex was outfitted with new tracking and timing systems, including the SuperFGD scintillator array and additional time projection chambers (TPCs), which provide finely segmented 3D tracking and precise particle identification~\cite{T2K:2021xwb,ND280Upgrade}. 
With J-PARC delivering more than 500~kW of beam power and upgrades planned toward 1.3~MW~\cite{OyamaJPARCUpgrade}, these improvements position T2K and its successor, Hyper-Kamiokande, to achieve leading sensitivity to $\delta_{\rm CP}$ and other neutrino oscillation parameters~\cite{T2K:2021xwb,ND280Upgrade,HyperKDesignReport}.  


\subsubsection{Neutrinos at the Main Injector (NuMI)}

The Neutrinos at the Main Injector (NuMI) beamline at Fermilab was the workhorse of the U.S. long-baseline neutrino program for nearly two decades. Beginning operation in 2005, NuMI enabled a sequence of landmark experiments: MINOS and MINOS+ provided precision measurements of atmospheric oscillation parameters and searched for sterile neutrinos; MINER$\nu$A made detailed cross-section measurements on a variety of nuclear targets; ArgoNeuT pioneered the use of liquid-argon technology on a neutrino beamline; and NO$\nu$A delivered world-leading results on neutrino mass ordering and the mixing angle $\theta_{23}$~\cite{Adamson:2015dkw,MINOS:2008foz}. Together, these results established Fermilab as the leading source of accelerator-based long-baseline neutrinos.

NuMI was originally conceived in the 1990s to serve MINOS, which required a controlled neutrino beam aimed 734~km toward the Soudan Underground Laboratory in Minnesota. Protons at 120~GeV from the Main Injector were directed onto a graphite target, and charged hadrons were focused by two magnetic horns before decaying in a 675~m atmospheric helium-filled tunnel. 
A unique feature of NuMI was its tunable design: by adjusting the relative position of the target and first horn, experimenters could shape the neutrino energy spectrum to match different physics goals~\cite{Kopp}. 
This flexibility allowed NuMI to evolve from the low-energy configuration optimized for MINOS to the medium-energy spectrum required for NO$\nu$A’s 810~km baseline to Ash River.

During its lifetime, NuMI underwent a series of upgrades that steadily increased beam intensity and reliability.
A major accelerator upgrade and beamline improvements expanded the facility's power to sustain megawatt-class operation from 400 kW at the start.
In 2024, NuMI reached a maximum hourly average of 1.018 MW, setting a record for long-baseline neutrino facilities.

The combination of NuMI's high power, flexible beam configuration, and near/far detector strategy made it an ideal platform for precision oscillation physics and neutrino interaction studies. 
This tunable, high-power design enabled precision measurements of atmospheric oscillation parameters, searches for sterile neutrinos, and detailed studies of neutrino interactions. This sets the stage for the physics goals of LBNF/DUNE.

\subsubsection{Booster Neutrino Beam (BNB)}

The Booster Neutrino Beam (BNB) at Fermilab has been a leading source of low-energy accelerator neutrinos for more than two decades. Operating in the sub-GeV to GeV range, it has supported a diverse program of experiments: MiniBooNE~\cite{MiniBooNE2009} and SciBooNE~\cite{SciBooNE2008} carried out cross-section measurements and sensitive sterile neutrino searches; MicroBooNE~\cite{MicroBooNE2019} pioneered precision liquid-argon measurements on a neutrino beamline; and today the BNB powers the Short-Baseline Neutrino (SBN) program, with ICARUS~\cite{ICARUS2021} and SBND~\cite{SBNDProposal}, as well as technology demonstrators such as ANNIE~\cite{ANNIEProposal}.  

The beam is produced by 8~GeV protons from the Fermilab Booster synchrotron, directed to a beryllium target to generate secondary pions and kaons. A single magnetic horn focuses the hadrons before they enter a short decay region, producing a neutrino spectrum that peaked near 800~MeV—an energy well matched to short-baseline oscillation studies. 
BNB can run in either neutrino or antineutrino mode by reversing the horn polarity, a flexibility that is valuable for sterile neutrino searches, precision cross-section measurements, studies of nuclear effects, and comparisons of oscillation behavior between neutrinos and antineutrinos.
Dedicated hadron production measurements by HARP have constrained the flux to within about 6–7\% in the peak region, unusually precise for a beam of this energy~\cite{BNB_FLUX,BNB_FLUX2}.  

The BNB has played a crucial role in neutrino physics. 
MiniBooNE provided some of the most stringent tests of the LSND anomaly, an excess of $\bar\nu_e$ events observed at Los Alamos and interpreted as a possible sterile-neutrino oscillation at the eV$^2$ scale~\cite{LSND1996}, 
while MicroBooNE demonstrated the capabilities of liquid-argon detectors for oscillation studies. 
A major goal of the SBN program is to definitively resolve the sterile neutrino question while providing detailed measurements of neutrino interactions at the GeV scale.
In parallel, the BNB has also been used to search for light dark matter and other exotic signatures, broadening its physics reach.  

BNB is expected to continue operating into the 2030s, supporting the complete physics program of the SBN and serving as a major facility for short-baseline oscillation studies, precision cross-section measurements, and searches for new physics.

\subsubsection{Long-Baseline Neutrino Facility (LBNF)}

Fermilab's Long-Baseline Neutrino Facility, now under construction, will provide the neutrino beam for the Deep Underground Neutrino Experiment, an international flagship in neutrino physics. 
Designed to send neutrinos 1300~km to the Sanford Underground Research Facility (SURF) in South Dakota, LBNF will deliver the most powerful accelerator-based neutrino beam to date. 
Together with DUNE’s massive far detectors, it will enable precision measurements of the CP-violating phase $\delta_{\rm CP}$, determine neutrino mass ordering, and test the three-flavor paradigm with unprecedented reach~\cite{lbnf-tdr}.  

The beam will be produced by 120~GeV protons from the Fermilab Main Injector, initially operating at 1.2~MW with a planned upgrade to 2.4~MW. Protons will strike a graphite target inserted into the first of three magnetic horns, which will focus secondary pions and kaons into a helium-filled decay pipe over 200~m long. Their decays will generate a wide-band muon-neutrino beam peaked in the 2–5~GeV range, optimal for studying oscillations at the 1300~km baseline. A hadron absorber and downstream muon monitors will provide control of residual flux and alignment.  

Several design advances distinguish LBNF from previous facilities. 
A three-horn focusing system expands the energy coverage and enhances CP sensitivity, while a helium-filled decay pipe reduces secondary interactions. 
The target hall and beamline incorporate extensive shielding, remote handling, and cooling systems to support sustained multi-megawatt operation.  

LBNF has been planned with scalability from the outset: the initial 1.2~MW beam will launch the DUNE physics program, while later upgrades will provide the exposures needed for high-precision studies of CP violation, mass ordering, and searches for new physics. 
As the successor to NuMI, it represents the next step in the long-baseline accelerator neutrino research.

\subsection{Future Directions}
Looking into the future, the HyperKamiokande experiment is expected to come online in the late 2020's, making use of the upgraded J-PARC neutrino beam.  Soon after, the LBNF beam will begin operations, delivering neutrinos to the DUNE experiment in the early 2030's.  Beyond that, the future of conventional horn-focused neutrino beams is less clear, but several concepts have been proposed.  
    \subsubsection{Upgrades to existing facilties} Essentially, all neutrino beams that have existed have gone through regular evolutions over time, with primary beams, focusing systems, and detector configurations morphing to accommodate new ideas.  The facilities described in ~\ref{sec:current} are no exception.  The NuMI beam will have run in a variety of configurations before its planned shutdown in 2028, after which components will be repurposed for the LBNF beamline.  The J-PARC neutrino beam will continue to run throughout the HyperKamiokande era, with upgrades to the J-PARC beam power and continual improvements to the near detector complex planned and other future evolutions under consideration.  The LBNF neutrino beam is expected to undergo modifications as it ramps up to 2+ MW in its first decade of operation.   Although not formally planned, LBNF could also be reconfigured in the future for alternate physics goals such as tau neutrino appearance and beam-dump like configurations for dark matter searches.  The BNB beam is currently approved to run through 2027 but its longer-term future remains under consideration.  
    Although it is not yet clear exactly what modifications these facilities will eventually undergo, it is clear that they will evolve.  

   Although this review does not cover pion decay-at-rest (piDAR) sources in detail, we note that the DAE$\delta$ALUS and IsoDAR concepts fall into this broader category. piDAR sources produce low-energy neutrinos from stopped pion decays and are closely related, both in source design and physics goals, to facilities such as SNS, J-PARC MLF, and ESS. Because of this connection and their relevance to future accelerator-based oscillation measurements, we briefly summarize DAE$\delta$ALUS
 and IsoDAR in the following.
   
    \subsubsection{DAE$\delta$ALUS} DAE$\delta$ALUS
 ~\cite{DAEDALUS_ISODAR} is a novel concept that proposes a pion decay at rest source that encompasses an $H^2+$ source, two cyclotrons, target station, and a beam dump.  The first cyclotron would accelerate the $H^2+$ ions to 60 MeV/amu and the second would accelerate them up to 800 MeV before they are sent through foils that strip the ions, yielding a beam of protons directed onto the target.  At these energies, pions produced in the target frequently stop within the target.  Negative pions are quickly absorbed by nuclei, but stopped positive pions decay, producing an isotropic source of primarily muon antineutrinos.  Neutrinos would be observed at a dectector at a baseline of 20 km.  As it is a fairly compact source, it could be built near a detector planned for another purpose, such as Hyper-Kamiokande.  The primary physics goal would be to measure CP violation via $\overline{\nu}_\mu\rightarrow\overline{\nu}_e$ oscillations.  
    \subsubsection{IsoDAR} IsoDAR~\cite{Isodar} is another potential decay-at-rest concept that would use a 60 MeV beam of protons directed onto a target composed of three nested beryllium spheres, which act as a Spallation neutron source. Neutrons are then captured by $^{7}\mathrm{Li}$ that surrounds the target. $^{8}\mathrm{Li}$ isotopes then decay, producing neutrinos with mean energy of 6.5 MeV.  The primary purpose of IsoDAR is a search for sterile neutrinos and other BSM physics.  Multiple locations for IsoDAR have been considered, the most recent being Yemilab in Korea, near the proposed Liquid Scintillator Counter (LSC) detector.  The cyclotron would need to provide 10 mA of current, substantially more current than is typically available in commercial cyclotrons.  However, mature designs of both the cyclotron and the neutrino production target have been developed.   
    \subsubsection{MOMENT} The Muon-decay Medium-baseline Neutrino Facility (MOMENT) experiment is a concept for a 300 MeV neutrino source produced with primary beam power of up to 15 MW.  Primary protons would be accelerated to $\mathcal{O}(1)$ GeV using a continuous wave linac similar to the China Accelerator-Driven System
(China-ADS) linac~\cite{ChinaADS}.   These protons would be directed onto a mercury jet target placed inside a high-field superconducting solenoid magnet.  Focused pions would decay in a short (tens of meters) pion decay region.  Muon decay products would be directed and charge separated by a series of magnets into a longer (hundreds of meters) muon decay volume, where muons would decay, producing a neutrino beam which would be observed by a detector 100-150 km away.  One possible setup would be for the beam facility to be at the China Spallation Neutron Source (CSNS) and the detector to be near the JUNO site, which are separated by approximately 150 km.  In order to definitively measure the CP violation, the muon-decayed MOMENT beam would provide advantages over planned pion-decay beams, such as a lack of any significant high-energy tail.  
However, the use of a 15~MW continuous-wave beam would pose substantial challenges, particularly in terms of radiation shielding, component activation, and target survivability.

    \subsubsection{ESSnuSB}  The European Spalation Source neutrino Super Beam (ESSnuSB)~\cite{ESSnuSB} proposes to double the current 5 MW power of the European Spalation Source (ESS) in Lund, Sweden, and to direct the new 5 MW of power towards a new neutrino facility composed of four target-horn systems.  This would produce neutrinos with energies between 0.2 and 0.6 GeV, which would be detected by a large water Cherenkov detector in the Zinkgruvan mine, approximately 360 km from the ESS.  This combination would provide the ability to measure the neutrino CP-violating phase with a precision of $8^\circ$.   ESSnuSB+ would add NuSTORM and ENUBET-like capabilities to the ESSnuSB concept.  NuSTORM is described in further detail below, and ENUBET was discussed in the previous section.  
    \subsubsection{NuSTORM} Neutrinos from Stored Muons, nuSTORM, ~\cite{nuSTORM} is a proposed muon storage ring neutrino source.  Although originally developed as a potential addition to the Fermilab accelerator complex, it has more recently been considered as a potential future project at CERN.  The CERN design would start with 100 GeV protons fast-extracted from the SPS and directed onto an inconel target fully contained within a focusing horn.  Focused pions would be transferred to the storage ring via a transfer line.  The storage ring is made up of two straights and two arcs.  Depending on their energy, up to 90\% of the pions decay in an initial production straight, and the resulting muons are energy selected and transported through an arc, a return straight, and a second arc back into the production straight.  A suite of detectors placed in line with the production straight would see flashes of neutrinos produced via muon decays in the production straight.  The primary physics motivations for NuStorm are precision measurements of neutrino interaction cross-sections enabled by a precisely measured neutrino flux,  searches for BSM physics.  Additionally, the facility would serve as a critical testbed for future neutrino factories and neutrino colliders.  
   
    \subsubsection{Neutrino Factories}  A Neutrino Factory is a step beyond the concept of a muon storage ring described above in the context of NuSTORM.  In this case, the intensity of the ring would be sufficient to enable a study of long-baseline neutrino oscillations.  Although this concept has existed since at least 1989~\cite{Geer}, there is currently no mature design for a neutrino factory.  However, recent phenomenological studies have considered the possibility of Neutrino Factories at Fermilab or BNL~/cite{Julia} and J-PARC~\cite{JNuFact}.  

     \subsubsection{ENUBET}
     The Enhanced NeUtrino BEams from kaon Tagging (ENUBET) project~\cite{ENUBET} is a novel concept for a monitored neutrino beam in which the flux is determined by directly observing charged leptons in the decay tunnel. Unlike most conventional beams, ENUBET does not rely on pulsed magnetic horns for focusing, but instead uses static focusing elements that allow for a slow and clean proton extraction and a well-collimated secondary beam. Instead of relying only on simulations of the beamline, ENUBET proposes watching the charged particles that accompany neutrino production inside the decay tunnel itself. The key idea is to detect positrons from so-called $K_{e3}$ decays, in which a charged kaon turns into a pion, a positron, and an electron neutrino.  Since every one of these decays produces a $\nu_e$, counting positrons provides a direct measurement of the electron-neutrino flux.  In the same way, muons from kaon and pion decays can be monitored to pin down the $\nu_\mu$ flux. To make this possible, ENUBET instruments the walls of the decay tunnel with a finely segmented sampling calorimeter.  This device measures how much energy particles deposit as they pass through layers of scintillator and absorber, and from the shower pattern it can tell positrons apart from muons and pions. By directly tagging the charged leptons, ENUBET can reduce the flux uncertainty to the percent level.  Such precision would allow neutrino cross-section measurements at the GeV scale that are limited not by beam knowledge but by the physics itself, and could directly benefit experiments like DUNE and Hyper-Kamiokande. After an intensive R\&D phase, including successful testbeam demonstrations at CERN, ENUBET is now being considered for a full implementation in the CERN North Area.  
     
\FloatBarrier
\section{CONCLUSIONS}
From the first demonstration of muon neutrinos at BNL in 1962 to today's megawatt-class facilities that drive oscillation measurements and neutrino interaction studies, accelerator-based neutrino beams have evolved. The concept of a high-energy proton beam, a meson production target, a magnetic focusing horn, decay tunnels, and downstream monitoring has remained basically the same, but advances in scale, precision, and reliability have made them more powerful. 
As past accelerator facilities established flavor and oscillation phenomena, current beams continue to extend these discoveries by searching for new physics. Together with Hyper-Kamiokande, LBNF and the DUNE experiment will provide unprecedented sensitivity to CP violation, mass ordering, and possible physics beyond the standard model in the future. In the decades to come, accelerator-based neutrino physics will remain central to the field due to their technological ingenuity and scientific reach. Looking ahead, ideas such as decay-at-rest sources, neutrino factories, and instrumented decay tunnels could open up new possibilities, keeping accelerator-based neutrino beams at the heart of discovery for many years to come.

\pagebreak

\section*{DISCLOSURE STATEMENT}
If the authors have nothing to disclose, the following statement will be used: The authors are not aware of any affiliations, memberships, funding, or financial holdings that
might be perceived as affecting the objectivity of this review. 

\section*{ACKNOWLEDGMENTS}
This material is based upon work supported by the U.S. Department of Energy, Office of Science, Office of High Energy Physics under Award Numbers DE-SC0023197 and DE-SC0022529. This work was also produced by Fermi Forward Discovery Group, LLC under Contract No. 89243024CSC000002 with the U.S. Department of Energy, Office of Science, Office of High Energy Physics.  The authors gratefully acknowledge R. M. Zwaska, with whom they developed the US Particle Accelerator School Neutrino Beams course on which this paper is loosely based.  We also thank Dan Cherdack, Fatima Abd Alrahman, Zarko Pavlovic, Jack Mirabito, and Pavel Snopok for assitance in locating neutrino flux simulations.   

%


\bibliographystyle{ar-style5} 
\bibliography{neutrinobeams} 

\end{document}